\documentclass[twocolumn,twoside,slac]{revtex4}
\usepackage{graphicx}
\usepackage{fancyhdr}

\newcommand{\figref}[1]{Figure~\ref{fig:#1}}

\newcommand{\figlabel}[1]{\label{fig:#1}}

\begin{document}

\title{A Web Based Document Database}

%

\author{E. W. Vaandering}
\affiliation{Vanderbilt University, Nashville, TN 37235, USA}

\begin{abstract}
 We describe a document database, developed for BTeV, which has now been
 adopted for use by other collaborations. A single web based database and
 archival system is used to maintain public and internal documents as well as
 documents for a related collaboration. The database performs archiving,
 versioning, access control, and easy remote access and submission. We cover
 the technical and security requirements of the database and its
 implementation. Usage patterns, improvements in our collaborative style, and
 missteps along the way are also discussed. 
\end{abstract}

\maketitle

\thispagestyle{fancy}


BTeV\footnote{URL: \url{http://www-btev.fnal.gov/}} is a Tevatron experiment
slated to begin taking data in 2008. We found ourselves in need of a system to
store and categorize documents throughout the life of the experiment (perhaps
until 2020). The software described here is designed to fulfill this goal for the
foreseeable future. While it may not last the entire lifetime of the
experiment, a transition to a new system should be reasonably easy. 

\section{Replacing our old document database}

Until the end of 2001, BTeV used a simpler document catalog. A few Perl scripts
and flat files were used to store a list of URLs and a small amount of information
about each of them. While this system worked, it was under-utilized and had
several problems and limitations.

The first problem was how documents were classified. Each document had
``topics'' (for instance a type of physics or a detector) that were
user-defined. This meant there was little consistency from user to another and
topics were effectively duplicated. Also, artificial limitations were placed on
documents; each document could have only one author, only one file, and the
lists of private and public documents were separate. These last two limitations
meant that documents were often duplicated. In the worst case scenario,
Postscript and PDF files would be provided for a private document which was
then released to the public, resulting in four ``documents'' even though the
content was the same.

Two other structural problems also made this old solution inadequate. First,
only URLs of documents were stored on a central server. The files themselves
remained on the Web at large, meaning that documents could (and did) easily
disappear. Finally, updated submissions overwrote the originals so there was
no way to keep a history of changes.

From inception in 1995 to decommisioning in 2001, this system was used to
categorize about 110 internal and 40 public documents. 

\section{Design of the new system}

In the process of replacing the old database, we wanted to eliminate all of
these problems and extend the database to deal with new situations as well. We
wanted a document database that would
\begin{itemize}
\item{Be single place to store and manage talks from collaboration meetings,
conference talks and proceedings, and publications \emph{and} to present the
relevant information for these
special cases.}
\item{Allow each document to have multiple revisions with old revisions still
available.}
\item{Allow each revision of each document to have multiple files. This
accommodates multiple file types (source and  presentable) and/or child files
(which is especially useful for web pages).}
\item{Provide the ability to upload files from the user's local computer or force the 
document database to download them via http.}
\end{itemize}

BTeV also had some security considerations not met by the document catalog. 
We have an associated group of computer scientists, BTeV
RTES (Real Time Embedded Systems\footnote{URL:
\url{http://www-btev.fnal.gov/public/hep/detector/rtes/index.shtml}}) which is
developing fault management software for the BTeV trigger system.  We want them
to be able to fully use the document database, but also to be able to have
documents which are not accessible to this group.  BTeV is also under active
review, so we needed the ability to easily provide certain documents to
reviewers.

To accomplish these tasks, we needed a database that
\begin{itemize}
\item{Had the ability to not only have public and private documents but also 
documents that are accessible to subgroups (like RTES or reviewers) with their
own password. We also want to have documents that are only 
accessible to sub-groups (like the executive council).}
\item{Could restrict reviewers to viewing selected documents, but not allow them
to create or
modify documents.}
\item{Provided the ability to easily move documents, or just certain versions of 
documents, among all these security settings.}
\item{Would restrict knowledge that a document exists to those that have 
privileges to view it.}
\end{itemize}

We evaluated the NUMINotes system\footnote{URL:
\url{http://www-numi.fnal.gov/noteSelPublic.html}} used by the MINOS
collaboration. While we found that it didn't solve most of the problems we were
having, its database structure and interface did provide good ideas for solving
our problems.

To implement our new document database (called DocDB) we use CGI
scripts written in Perl. We choose MySQL as the relational database since Perl
bindings are readily available and its (somewhat limited) features were more
than adequate for our needs.

The files that make up documents are stored on the web server's file system.
All the meta-data for the documents are stored in a collection of MySQL tables
shown in \figref{db_model}. This table structure has allowed us to remove the
arbitrary limits on numbers of authors, topics, and files per document. For
instance, as shown in \figref{db_model}, revision-author pairings are stored in
a separate table and an arbitrary number of these pairings can be associated
with each document revision and with each author. Linking between tables is
supported with a number of keys.

\begin{figure*}
\includegraphics[width=0.75\textwidth]{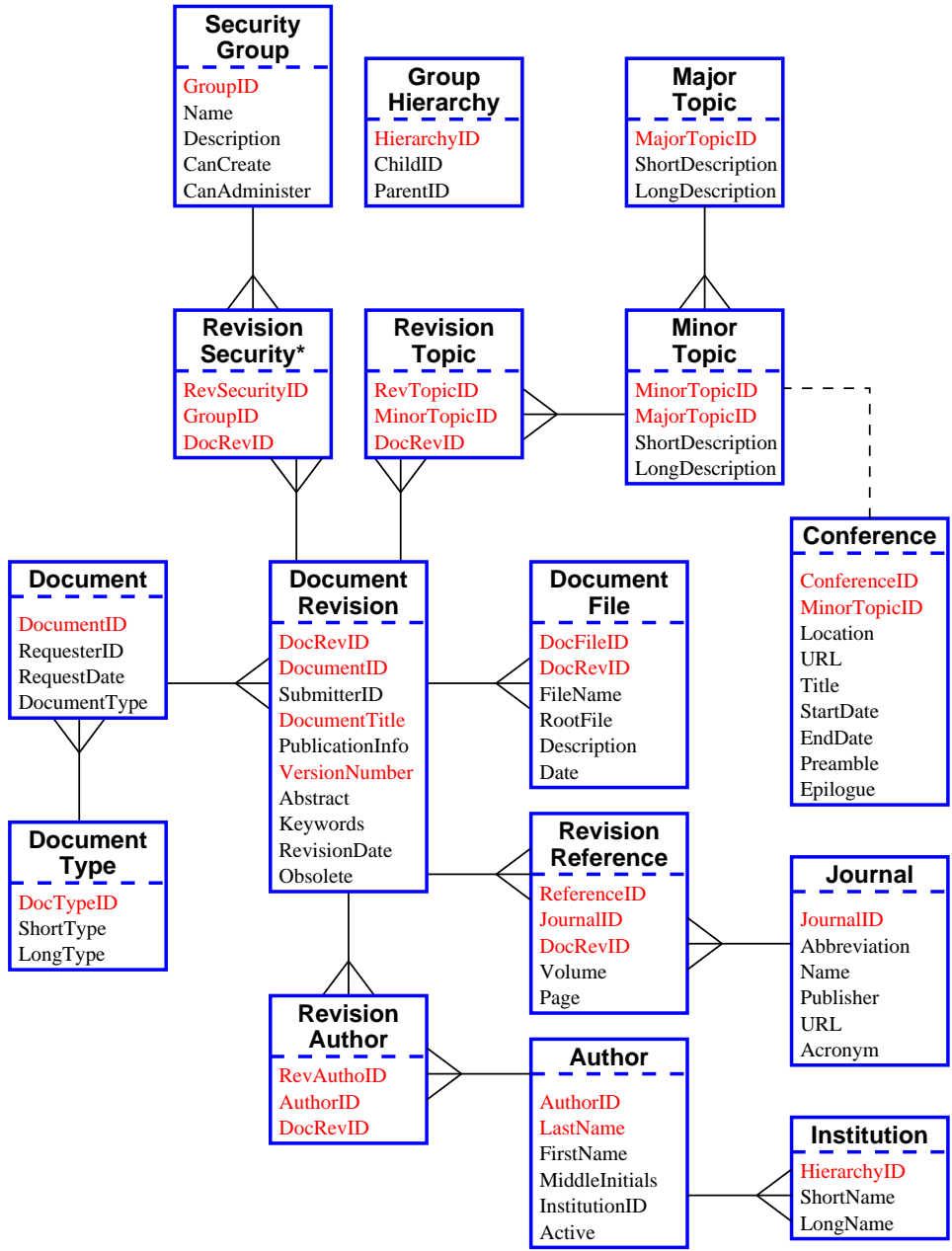}
\caption{Database table structure of the DocDB. Fields in red (or light
color) are indexed for fast lookup. In addition to
RevisionSecurity there is a parallel table named RevisionModify which, in the
optional enhanced security model, contains the list of groups allowed to modify
documents. See the text for more details.}
\figlabel{db_model}
\end{figure*}

These different tables are, of course, hidden from the user. When asked for
information on a certain document, the information is gathered from a number
of different tables (which can be up to four ``steps'' away from the original
table) and all the information is presented in a single summary. An example of
such a summary is shown in \figref{show_example}. 

\begin{figure*}
\includegraphics[angle=-90,width=\textwidth]{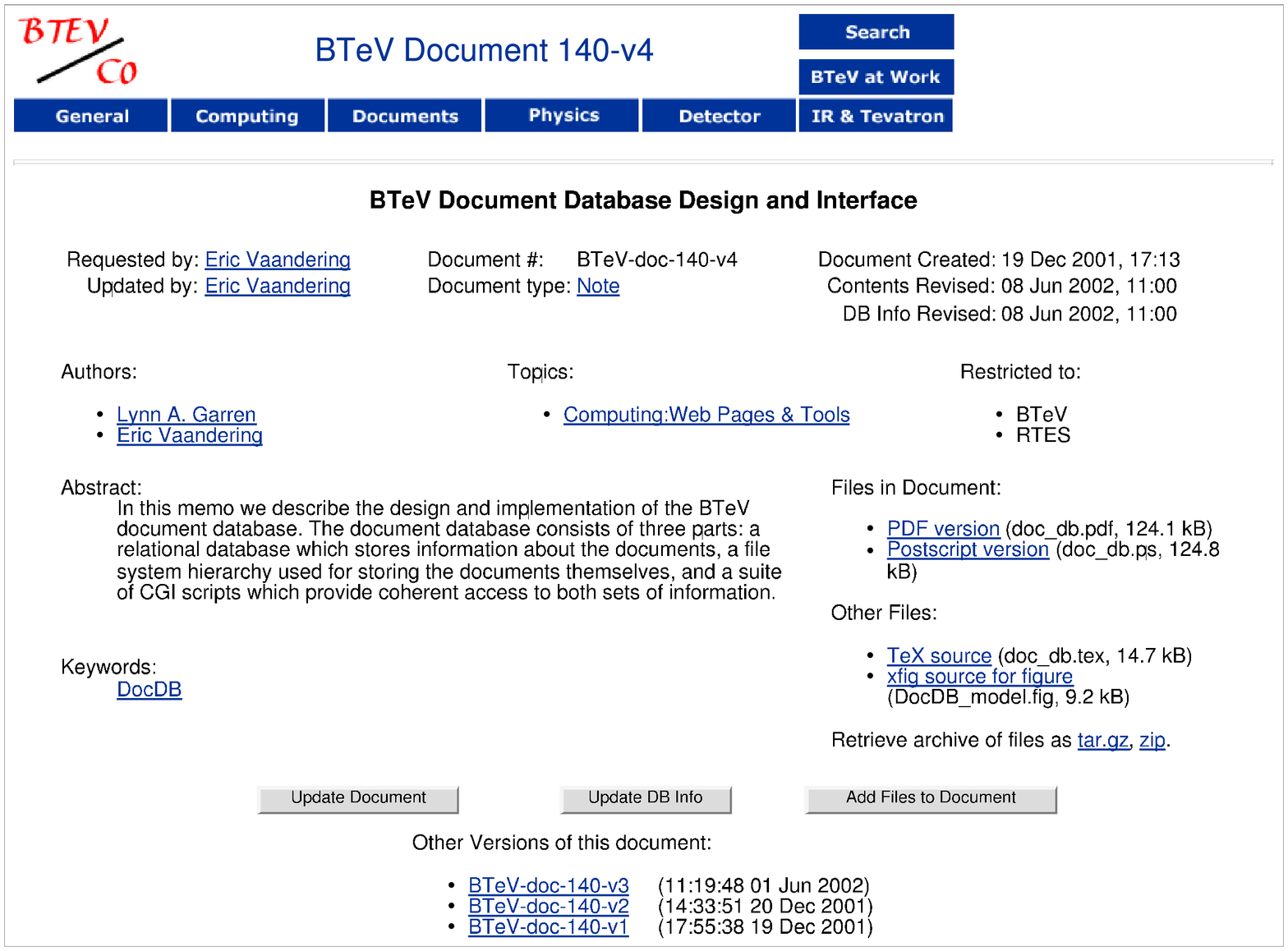}
\caption{An example document view. See the text for an explanation.}
\figlabel{show_example}
\end{figure*}

\subsection{Adding documents}

Users can add or modify information in a number of ways. They can reserve a
document number for a future document (for reference purposes). They can create
a new document, a new version of an existing document, or just update the
document meta-information without supplying new content.  They can also add or
replace files (perhaps a new presentation format) without creating a new
version. 

Most of the form used to submit documents is shown in \figref{enter_example}.
At the top, the user is asked to enter a title, abstract, and any keywords
which describe the document. Next is the box (or boxes) to upload the files in
the document. In an alternate version, the user may request that the DocDB
fetch a file via http or ftp from a server (which may be password protected). 

Next, the author must choose a document type, select who he or she is
(Requester), who all the authors of the document are, and what groups may view
the document. Finally, the user selects all the subtopics which apply to the
document. (Here there is one list of subtopics for each topic).

Not shown are the elements to specify a revision date, journal references, and
a field to enter other publication information for the document. The
appearance of this form is customized based on user preferences and requests;
for instance, a user can ask to always have upload boxes for three files. What
is shown here is the simplest possible version.

\begin{figure*}
\includegraphics[width=\textwidth]{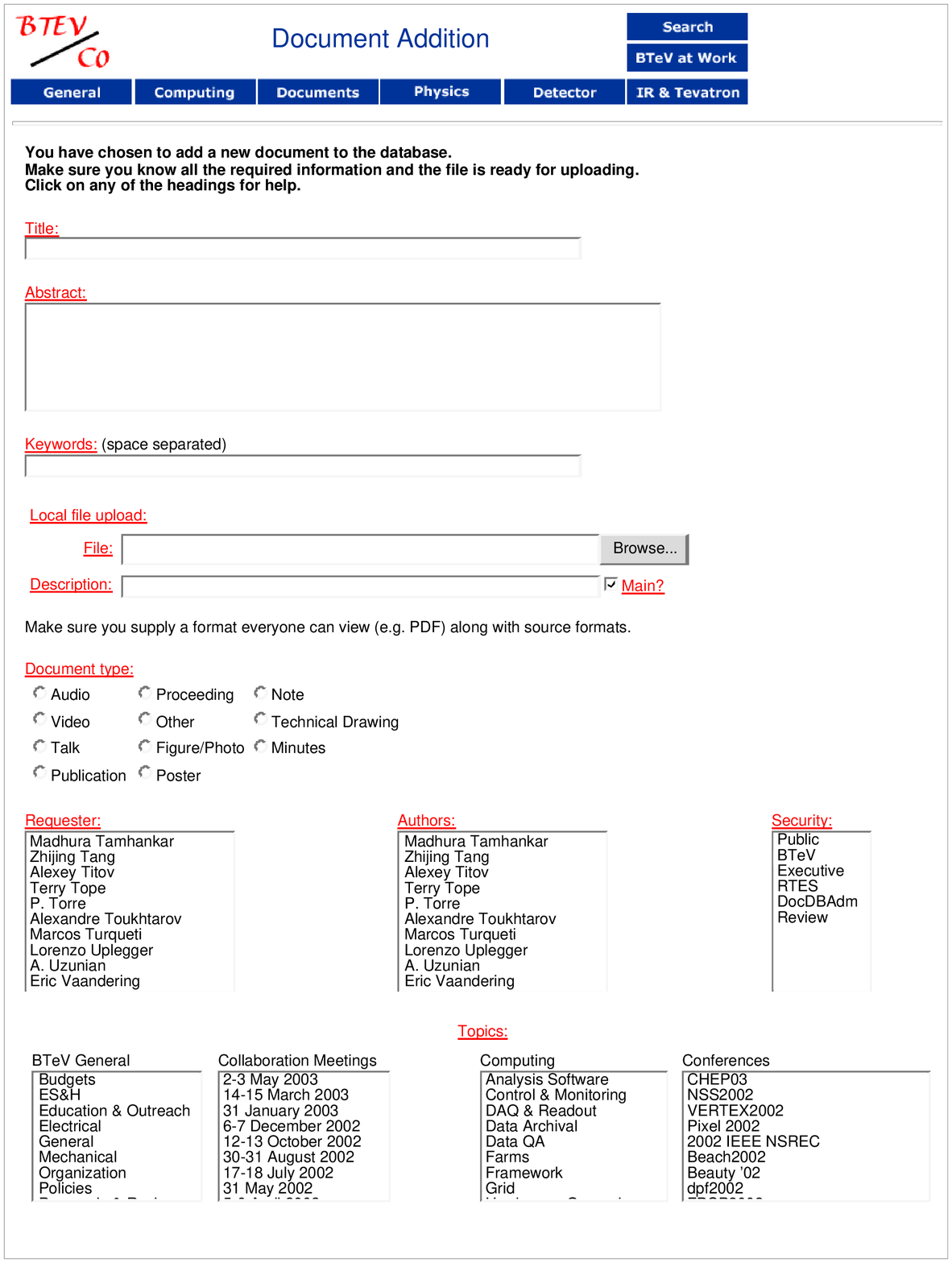}
\caption{Form for entering a document. Now shown are all the topics, references,
and the area to enter additional publication information. See the text for a
full description.}
\figlabel{enter_example}
\end{figure*}

\subsection{Document classification}

The document view in \figref{show_example} shows many of the different ways in
which documents can be classified. For example, at the top left, we see the
individual responsible for putting the information into the database and the
authors of the document. At  top middle we see the document ID
(BTeV-doc-140-v4) and document type. Underlined text are hyperlinks to a list
of documents with, for example, the same author. In the center are the
subtopics for the document (explained below). Near the bottom right are links
to the actual files in the document; some are designated as primary files and
some are designated as supporting files. Near the bottom left are the user
supplied keywords for the document and at the bottom center are links to
earlier versions of the document. 

There are also buttons that the user can use to add a new version, change the
meta-information, or add files to this document.

\textbf{Topics: }Because of our earlier, confusing, experience with allowing
users to freely define new topics to classify documents, we adopted a
centralized approach. We have about ten topics (MajorTopics in the internal
nomenclature) each of which is divided into subtopics (MinorTopics internally).
These lists can only be modified by the administrators of the database. For
instance, ``Detectors'' is a topic and sub-detectors like ``Pixel Detector''
are subtopics. For detector subtopics, each is used primarily by a different
subgroup of the collaboration numbering about 10--20 people. 

\textbf{Keywords: }
In order to give these subgroups the ability to better organize their own
documents, we introduced the concept of keywords, which are free form and 
entered by users.  Subgroups are encouraged to form canonical lists of
these keywords to be affixed to the relevant documents, so for instance, the
pixel detector group may choose the keyword ``support'' to classify any
documents dealing with the pixel support structure.

\textbf{Authors: }
All possible authors are also maintained in a managed list. While there was
some initial concern about the flexibility of this, it has worked out well.
Duplication and consistency issues, as with topics, were the reason for this
choice.

\subsection{Security}

In order to achieve our security goals, we have implemented a parent/child
security model. Each group has its own password, and parent groups may view any
document their child groups may. The security mechanism is based on http Basic
Realm Authorization. The meta-information is protected by the Perl scripts, the
actual files by \texttt{.htaccess} files; which security method is protecting
what is mostly transparent to the user. In the normal mode of operation, any group
that may view a document may modify that document (presuming they can modify
documents at all), but there is an alternate security model in which the groups
that may modify a document are separate from the groups that may view it.

\section{Special cases}

In addition to storing general documents as they are produced, we realized that
there are certain special types of documents which need additional information
stored about them, and/or need to be presented to the user in special ways.

\subsection{Conferences and collaboration meetings}

Users associate their documents with a conference or meeting by selecting a
special topic in addition to the topic(s) of the document. These topics have
additional information associated with them, such as dates, location, and a URL
for the conference or meeting.

BTeV has collaboration meetings about every six weeks. These are video
conferenced, so an accurate and timely record of the meeting is important for
remote participants. Since uploading a talk for a meeting is many people's
first interaction with the DocDB, a special form to enter such talks into the
database was created. This minimizes the information that a user must fill in
and has sensible defaults for group meeting talks. However, users can always
use the full-featured entry form as well. 

There is also a special default mode to view the list of talks. The files
(PowerPoint, PDF, etc.) for the talk are linked directly from the list of
talks, since this list of talks is used as the ``home page'' for the meeting
while it is going on.  For conferences, there is also a special view of the
document list.

\subsection{Publications}

Published documents also receive special treatment. When placing a document in
the database, publication references for that document can be entered. There
may be an arbitrary number of such references for each document (e.g. an
arXiv.org and journal reference). There is also an extensible framework to
generate external links to the paper on the journal's website.  

\section{Other Features}

The most used parts of the system are those allowing addition of documents,
showing individual documents, and retrieving lists of documents by an
author, etc. However, there are several other features of the DocDB that,
while less frequently used, provide needed functionality.  

\subsection{Searching}

We have provided a robust search engine that allows searching for individual
words, phrases, etc. in all of the text fields and  searches on authors,
topics, and the other fields in the database. For a database of our size,
all the flexibility provided isn't strictly necessary, but as the number
of documents increases it will become more important. Also, listing by
keyword uses the search engine to find the relevant documents.

\subsection{Preferences}

The DocDB has lots of options, especially regarding how the document
submission page appears. For instance, a user can upload from a local file or
submit a document via http. We have provided a preference system, based on
browser cookies, to save the user's preferences for these options.

\subsection{E-mail notification}

We have also provided the ability for the DocDB to send e-mail to users when
documents they are interested in have been added or modified in the database.
Users can select to be notified based on document topics, keywords, or authors,
and they can select to be notified about such changes immediately, nightly, or
weekly. They can have separate criteria for each time
frame.

\subsection{Administration interface}

Finally, of interest primarily to the administrators of the document database,
is a set of tools which allow the administrators to modify the lists of
authors, topics, document types, etc. This means that once the database is set
up, the administrator should never need to manipulate the underlying SQL database.
Everything that needs to be done can be done via the web interface.

For convenience, certain administration functions are left open to all users,
such as adding a person to the list of potential authors and adding a
conference.

\section{Effects on collaborative style}

We have generally found the DocDB to be a great benefit in enhancing
collaboration within BTeV. In this section, we will give some examples of
these benefits.

First, we have been fairly effective in redefining  what a ``document'' is. A
document, for us, is not just words written down on a piece of paper, but is
any information which someone wants to save and share with others. Users also
seem to have accepted a lower threshold of ``importance'' that they feel
information must meet before placing information in the DocDB. While most
things in the DocDB still fit a traditional definition of documents, we have a
number of photograph collections, videos, musings, and figures which have been
placed in the DocDB. 

We've also been fairly effective in promoting the DocDB as a central
repository of information. In the past, most sub-groups (mostly sub-detector
projects during this stage of BTeV) maintained, on web pages, lists of various documents.
With an easy to use and visible DocDB, more of this information is now being
placed in the database. However, migration of  older information into
the DocDB has been slow or non-existent for most groups.   

One application that has benefitted most strikingly from using the DocDB is our
video conferenced collaboration meetings. In the past we had an agenda with
links to talks maintained by a secretary. This caused several problems. First,
meetings were often held on weekends which meant updates weren't made. Even
when the agenda was updated, there was a delay before a submission appeared
on the list. What this often led to was a flurry of e-mails, sometimes with
large attachments, as a person got up to give their talk. The speaker would
then be delayed while all the remote groups checked their e-mail and accessed
the talk. 

Now the process is much smoother. Speakers can easily update a talk just
moments before giving it, often incorporating information or ideas from
earlier talks. Additionally, talks are archived in the same location as other
documents. 

As mentioned above, to further ease the process of posting and viewing talks,
slight modifications to both the submission and listing interfaces have been
made.

External reviews of the experiment are another place where the DocDB has
proved its worth. We have a read-only account for reviewers and relevant
documents are easily made accessible to them. The group organizing
our reviews has been especially effective in using keywords to further
organize review documents. Reuse of information is another benefit; our
latest review (in Fall 2002) required over 250 documents, but many of these
will simply be updated for future reviews.

As mentioned above, our old database was used to catalog about 150 documents
over a 7 year period from 1995--2001. From the end of 2001 until
mid-2003 we've placed over 1750 documents in the new database. Of these,
about 400 documents actually predate the DocDB. Our collaboration
(about 200 people including engineers, etc.) averages about 3--4 new documents
placed in the database per day. The number of ``living documents,'' or older
documents which are regularly updated, remains fairly low.

As mentioned above, our old document catalog enforced lots of artificial
limitations on document classification. Currently the average document has 1.6
versions. Each version averages 1.4 authors, 1.8 topics, and 1.8 files,
confirming that these earlier limitations were not valid.

\section{Other users}

While the DocDB was initially designed for BTeV and not much thought was put
into a portable system, other groups have been evaluating or have started to use the
code. The underlying Perl code is now completely portable and setting up the
DocDB takes just a couple of hours. 

Fermilab's Beams Division is the only other user so far to officially adopt the
DocDB and a small number of features were added to the code to facilitate
their use. Their use began with the Main Injector group, but the system was
quickly adopted by the rest of the division. 

The Beams division began using this system at the beginning of 2003 and, as of
this writing, has nearly 600 documents in their DocDB. Their usage patterns seem
to be similar to BTeV's. 

Like BTeV, they have found the DocDB to be very useful in organizing their
frequent reviews and meetings. Previously most of their meeting
talks existed in hard copy only, archived in a filing cabinet. However, they
now find it just as easy to make an electronic copy and place it in the
database before the meeting. Beams also seems to be more effective than BTeV in
using keywords to better organize their documents. Their database
(\url{http://beamdocs.fnal.gov/cgi-bin/public/ DocDB/DocumentDatabase}) is a
particularly good demonstration of the software since many of the documents are
publicly accessible.

The DocDB code and installation instructions are available for use by other
groups at the URL \url{http://cepa.fnal.gov/DocDB/doc/install-docdb.html}.  A
working demo  of the software will be provided at this address as well, perhaps
by the time of publication. We welcome additional users and are willing to
provide some help setting up the system. We are also open to extending the DocDB
to meet others' needs.

\section{Lessons learned}

While we've been very happy with this system, we've also made a few mistakes
along the way and learned some lessons. 

The first thing we've learned (or confirmed) is that effectively organizing
groups of physicists can be hard. We've never had any ``official'' guidance on
using the DocDB on a collaboration wide basis, we've only had suggestions on how
to go about using the DocDB and organizing documents within it. Some subgroups,
often consisting of larger numbers of engineers, have provided such official
guidance and they use the database in a more consistent and effective manner.
(In fact, the concept of keywords was added to the DocDB o accommodate these
groups).

The second thing we've learned is that limiting flexibility for fear of misuse was
not a good idea. Initially we resisted allowing as much flexibility as we have
now in creating or modifying documents because that capability can be used by a
knowledgeable (or particularly n\"{a}ive) user to circumvent the versioning and
archiving features of the database. However, this has not proved to be a problem
as users overwhelmingly use this power in a responsible fashion.

Finally, we've learned from our seven year experience with our old catalog
system that documents do disappear from the web, that archiving is very useful,
and that starting with a flexible system as soon as possible is very helpful.
However, we've also learned that the initial pain of migration from an
inadequate system is well worth the benefits.

\section{Conclusions}

In conclusion, BTeV has built a new document database an has now been using and
refining it for a year and a half. This new system is much more flexible than
our previous one and
the other solutions we looked at. This new software has made noticeable
improvements in the way we collaborate and has been a real time saver.

We've also shown that this software is easily adaptable to other groups and
that they are reaping similar benefits.



\end{document}